\documentclass[aps,preprint,showpacs]{revtex4}

\usepackage{graphicx}
\usepackage{amsmath}

\def \G{\mathcal{G}}

\begin{document}

\title{Liquid-gas phase transition in nuclear matter from realistic many-body approaches}

\date{\today}

\author{A. Rios}
\affiliation{National Superconducting Cyclotron Laboratory and
Department of Physics and Astronomy, Michigan State University,
East Lansing, 48824-1321 Michigan, USA}
\email[]{rios@nscl.msu.edu}

\author{A. Polls and A. Ramos}
\affiliation{Departament d'Estructura i Constituents de la Mat\`eria and Institut de Ci{\`e}ncies del Cosmos, 
         Universitat de Barcelona, Avda. Diagonal 647, E-08028 Barcelona, Spain}

\author{H. M\"uther}
\affiliation{Institut f\"ur Theoretische Physik,
             Universit\"at T\"ubingen, 
             D-72076 T\"ubingen, Germany}

\begin{abstract}
The existence of a liquid-gas phase transition for hot nuclear systems at 
subsaturation densities is a well established prediction of finite 
temperature nuclear many-body theory. In this paper, we discuss for the first time
the properties of such phase transition for homogeneous nuclear matter within 
the Self-Consistent Green's Functions approach.
We find a substantial decrease of the critical temperature with respect to 
the Brueckner-Hartree-Fock approximation.
Even within the same approximation, the use of two different 
realistic nucleon-nucleon interactions gives rise to large differences in the
properties of the critical point.
\end{abstract}


\maketitle

\section{Introduction}
\label{sec:intro}

The study of the thermal properties of nuclear matter in terms of realistic many-body 
approaches has received little attention in the literature, in spite of its 
potential applications in astrophysics and in the physics of heavy ion collisions.
Dense hadronic matter appears in some astrophysical scenarios and it can be particularly 
hot in the very early stages after the creation of a proto-neutron star in a type-II supernova explosion. 
Typical temperatures for such systems are in the range of $20-60$ MeV \cite{prakash96}.
It is possible that some astrophysical observables are influenced by the 
presence of such high temperatures, because of the modifications induced to the bulk
and microscopic properties of dense matter.
As an example, the gravitational wave spectrum generated in a neutron star merger might be
influenced by the temperature dependence of the equation of state \cite{janka07}. 
Also, the cooling of the neutron star after its birth is dominated by neutrino emission. This
involves a series of processes which are particularly sensitive to the formation of Cooper 
pairs in the nuclear medium  \cite{yakovlev04}. 
Consequently, astrophysical observations could help us to constrain the temperature dependence of the microscopic and bulk properties of dense matter. 

There is however another way to access the thermodynamical (TD) properties of nuclear and hadronic systems.
``Hot'' nuclei are created in experimental facilities, in the collisions of heavy ions
at intermediate energies \cite{pochodzalla97}. There
has been an increasing effort to interpret experimental data concerning these heavy-ion collisions
in terms of the equation of state of nuclear matter \cite{siemens83,dani02}. More specifically, 
multifragmentation reactions at intermediate energies  are used to access
the properties of thermally equilibrated ``blobs'' of nuclear matter \cite{das05}. The 
evidence gathered in these experiments points towards the existence of a 
liquid-gas phase transition for nuclear systems at densities below the empirical
 saturation density, $\rho_0 = 0.16$ fm$^{-3}$, and temperatures around $\sim 6-9$ MeV \cite{pochodzalla95}. 
Empirically, the liquid-gas phase transition is usually discussed in terms of a plateau in the
 caloric curve for different types of reactions at different energies 
\cite{natowitz02}, although the presence of such plateau may also be explained in 
terms of a density and temperature dependent effective mass \cite{sobotka04}.  
Statistical models, which
assume an equilibrated thermal freeze-out density, have had certain success in 
describing experimental results \cite{tsang06}.
In any case, a common underlying idea in all these discussions
is the assumption that some sort of thermal equilibrium is 
reached at a given stage of the multifragmentation reaction. 

A first step towards a full theoretical understanding of the thermal properties of nuclear systems
can be achieved by studying symmetric nuclear matter (an ideal, infinite, high-density system
 composed of the same amount of neutrons and protons interacting via the strong force) at
finite temperatures, because the features of 
the liquid-gas phase transition are in general easier to study in the homogeneous
system. 
This is obviously a very crude
approximation and can only be taken as a guide for a better theoretical understanding of the thermal properties of dense matter. As an example, the low-density phase of 
dense nuclear matter is not a homogeneous gas of nucleons and should instead 
be described in terms of droplets of light nuclei (deuterons, alpha particles).
In the case of real experiments, moreover, finite size effects are as
important as the bulk properties in determining the critical behavior. 
Finite nuclei can only be excited up to some limiting temperature above which
Coulomb effects together with the decrease in surface tension lead to their thermal dissociation\cite{levit85,baldo04}.  All in all, it is not an easy task to link the
properties of the phase transition in  the homogeneous case to those 
of finite nuclei \cite{jaqaman83,kapusta84,goodman84}. 
In this paper, however, we shall concentrate on studying the
liquid-gas phase transition in the ideal case of nuclear matter. Our basic goal is to 
discuss to what extent the properties of this transition depend on the many-body description
chosen as well as on the underlying NN interaction.

Traditionally, the studies of nuclear systems at finite temperature have been performed 
using effective interactions within mean-field theories, either relativistic 
\cite{muller95,hua00} or non-relativistic \cite{sil04,cho03}.
The temperature dependence in these approaches arise mainly from the modification
of the zero temperature step-like momentum distributions, which become Fermi-Dirac distributions. 
As a consequence, the mean-field and the bulk properties have a rather simple dependence
on temperature. Thermal effects on the correlations between the
strongly interacting nucleons are completely ignored.
The effective interactions, which are fitted to describe the bulk properties of
nuclei at zero temperature and thereby account for correlation effects
in a phenomenological way, have no temperature dependence. This is
in striking contrast with the more microscopically founded many-body 
calculations \cite{lejeune86,baldo99,rios05}, where the in-medium interaction is found
by using some sort of Pauli blocking. Such blocking effects are weakened by temperature, 
therefore giving rise to non-trivial temperature dependences in the microscopic
and macroscopic properties of dense matter.

There are only a few realistic many-body calculations for nuclear matter at finite temperature
\cite{friedman81,baldo88,bozek06}. Some of these have consistency problems problems, since they 
are not truly based on finite temperature many-body theory, but on naive extensions of 
zero temperature approaches to the non-zero temperature domain. The Green's functions approach is however based on the perturbative expansion of the single-particle propagator at finite temperature \cite{fetter} and therefore it is a well-grounded approach which allows for systematic improvement.
In contrast, the variational approach is based on an explicit incorporation of the two-body 
nucleon-nucleon (NN) correlations
in the nuclear wave function. Its extension to finite temperatures has traditionally relied on the ``frozen'' 
correlation approximation, \emph{i.e.} the correlation functions have been assumed to be the same 
at zero and at finite temperature \cite{friedman81,moshfegh05,kanzawa07}. 
This might be adequate for the high-density low-temperature
phase, but it remains to be seen if it offers an appropriate description of the high-temperature phase. 
Recently there has been a substantial effort to discuss the more
formal aspects associated to the variational approach at finite temperatures \cite{mukherjee07}. 
Let us also note that the variational approach is restricted by nature to deal with local potentials, while the SCGF can in principle deal with any sort of realistic two-body force. 
The Brueckner--Hartree--Fock (BHF) 
approach has also been used in the study of hot nuclear systems \cite{baldo88}, although
the Brueckner--Bethe--Goldstone expansion, in which the
BHF approach is based, is only valid at zero temperature. The standard finite temperature 
generalization of this approach is, in a way, phenomenological and relies
on the replacement of all the 
step-function momentum distributions of the zero temperature case with Fermi-Dirac ones.
This finite temperature extension is however not
well-defined at a fundamental level, since it does not take into account the 
contributions of anomalous diagrams \cite{luttinger60a}. 
A consistent Brueckner--like approach at finite temperature is given by the Bloch-de 
Dominicis formalism \cite{bloch58b,bloch59a,bloch59b}, which has only recently been 
applied to the nuclear many-body problem at finite temperatures by Baldo and coworkers \cite{baldo99,baldo04}. 
Relativistic effects have also been explored in the framework of 
an extension to finite temperature of the Dirac-Brueckner-Hartree-Fock approach \cite{terhaar86,huber98}.

A good alternative for a suitable microscopic many-body description of hot correlated systems
is provided by the perturbation expansion of the one-body Green's functions 
\cite{kadanoff62}, which relies on the generalization
of the Wick theorem at finite temperature \cite{mattuck}. Due to the strong short-range repulsion of the  
NN interaction, the minimum meaningful scheme which can describe nuclear matter
is provided by the ladder approximation.
This arises from a decoupling of the three-body Green's function in terms of one- and two-body propagators
\cite{mattuck} and can be cast as a set of self-consistent equations that describe the in-medium
modifications of the nucleon due to the presence of the surrounding nucleons. These equations 
lead to an approximation that goes beyond the mean-field and the quasi-particle pictures,
 \emph{i.e.} the off-shell effects and the fragmentation of single-particle states are fully
taken into account. This Self-Consistent Green's Functions (SCGF) approach is well established in
nuclear physics 
\cite{dickhoff04} and has already been applied to study the microscopic properties of nuclear matter 
at finite temperatures \cite{alm96b,frick03,frick05,bozek1}. A major motivation for these 
studies has been the fact that, at non-zero temperature, one can avoid the numerical 
and physical problems associated to the 
neutron-proton pairing instability \cite{vonder93,bozek99,dickhoff05}. 
The SCGF, however, can also be used to study the TD properties
of the system in the normal phase, by making use of the Luttinger-Ward (LW)
formalism. This approach leads to thermodynamically consistent results, once the effects
of correlations in the entropy have been carefully taken into account \cite{rios06,riosphd,bozek06}.
In the following we shall use this formalism to study the properties of the liquid-gas phase transition
in nuclear matter. Note however that the effect of three-body forces is not included in our approach and, as a consequence, the saturation properties of nuclear matter are not well reproduced. The lack of three-body forces will also have an impact on the liquid branch of the phase transition, thus modifying to a certain extent our predictions for the critical properties of nuclear matter. Our results should be considered as a theoretical study quantifying the importance of short-range correlations on the liquid-gas phase transition. 

In the next section, we briefly summarize the SCGF approach at finite temperature and we 
discuss under which approximations the standard BHF at finite temperatures can be obtained from it. 
The third section will deal with the application of the LW formalism to the calculations
of the TD properties of a correlated system of nucleons. 
The results for the liquid-gas phase transition will be discussed 
in the fourth section. Finally, a brief summary will be given in Section V.

\section{SCGF at finite temperature}

The key quantity in many-body Green's functions theory is 
the single-particle propagator which, 
in the grand-canonical ensemble, is defined according to:
\begin{equation}
i \G({\bf k}t, {\bf k'}t') = \textrm{Tr} \left \{ \hat \rho \mathcal{T} [ a_{\bf k}(t) a_{\bf k'}^{\dagger}(t')] \right \} \, ,
\label{eq:green1}
\end{equation}
where we have introduced the density matrix operator:
\begin{equation}
\hat \rho = \frac {1}{Z} e^{- \beta (\hat H - \mu \hat N)} \, ,
\end{equation}
and the partition function:
\begin{equation}
Z= \textrm{Tr} \left \{ e^{-\beta ( \hat H - \mu \hat N)} \right \} \, .
\end{equation}
In the previous equations, $\beta=1/T$ denotes the inverse temperature and $\mu$
is the chemical potential. $\mathcal{T}$ stands for
a time-ordering operator and the traces $\textrm{Tr}\{ \cdot \}$ are to be taken over
all energy and particle number eigenstates. 
One can express the single-particle propagator in Fourier-energy space
by means of the spectral decomposition:
\begin{equation}
\G(k,\omega) =  \int_{-\infty}^{\infty} \frac {\textrm{d} \omega'}{2 \pi}
\frac {A^{>}(k,\omega')}{\omega - \omega' + i \eta} +
\int_{-\infty}^{\infty} \frac {\textrm{d} \omega'}{2\pi} 
\frac {A^{<}(k,\omega')}{\omega - \omega' - i \eta} \, ,
\end{equation}
where the finite-temperature equivalent of the $T=0$ hole spectral
function is given by the Lehmann representation:
\begin{equation}
A^{<}(k,\omega) = 2 \pi \sum_{n,m} \frac {e^{-\beta (E_m-\mu N_m)}}{Z}
\mid \langle \Psi_n \mid a_k \mid \Psi_m \rangle \mid^2 \delta \left [\omega
-(E_m -E_n)\right ] \, .
\end{equation}
The main difference with respect to the zero temperature case comes from the
average over the thermal bath in the initial states.
A similar definition holds for $A^{>}(k,\omega)$, with the replacement 
$a_k \to a_k^{\dagger}$. In thermal equilibrium, both spectral functions
are related by the Kubo-Martin-Schwinger relation:
\begin{equation}
A^{>}(k,\omega) = e^{\beta (\omega -\mu)} A^{<}(k,\omega) \, .
\end{equation}
In contrast to the zero temperature case, 
the energy domains of $A^{<}(k,\omega)$ and $A^{>}(k,\omega)$ are not separated 
by the Fermi energy and both spectral functions are defined for all energies. 
The total spectral function, $A(k,\omega)$,
is given by the sum of the two functions, $A^{<}$ and $A^{>}$,
and therefore it can be expressed in 
terms of the values of $\G$ close to the real axis:
\begin{eqnarray}
A(k,\omega) &=& -2  \textrm{Im}  \G(k,\omega_+)        \, 
\label{eq:asf}
\end{eqnarray}
(where we have introduced the notation $\omega_+=\omega+i \eta$).
Since the spectral function completely determines the one-body propagator via
the previous equation, all the one-body properties of the system can be expressed
in terms of it. 

In the medium, the single-particle Green's function can be obtained from Dyson's equation:
\begin{eqnarray}
\bigg[ \omega - \frac {\hbar^2 k^2}{2 m} - \Sigma(k,\omega) \bigg] \G(k,\omega) &=& 1 \, ,
\label{eq:dyson}
\end{eqnarray}
where $\Sigma(k,\omega)$ denotes a complex self-energy. 
The self-energy accounts for the interactions of a particle with the other particles in the 
medium. It fulfills the following spectral decomposition:
\begin{eqnarray}
\Sigma(k,z) &=& \Sigma^{HF}(k) - \int \frac {\textrm{d} \omega}{2 \pi} 
\frac {2 \textrm{Im} \Sigma(k,\omega_+)}{z-\omega}        \, ,
\label{eq:sigma_dec}
\end{eqnarray}
where $z$ is a complex variable and the term $\Sigma^{HF}(k)$ is a real energy-independent
generalized Hartree-Fock contribution:
\begin{eqnarray}
\Sigma^{HF}(k) = \int \frac{\textrm{d}^3 k'}{(2\pi)^3}
\langle \mathbf{k k'} | V | \mathbf{k k'} \rangle_A \, n(k') \, ,
\label{eq:HF_self}
\end{eqnarray}
with $n(k)$ being the single-particle momentum distribution:
\begin{eqnarray}
n(k) = \int \frac{\textrm{d} \omega}{2 \pi}  A(k,\omega) f(\omega) \, ,
\label{eq:nk}
\end{eqnarray}
and where $f(\omega) = \left[ e^{\beta (\omega - \mu)} + 1 \right]^{-1}$ stands for the 
Fermi-Dirac distribution. 
The imaginary part of the self-energy, necessary to compute the second term in Eq.~(\ref{eq:sigma_dec}),
is obtained by letting $z \to \omega_+$ in Eq.~(\ref{eq:sigma_dec}) and it is related to 
the effective two-body NN interaction in the medium
(the so-called scattering $T$-matrix):
\begin{align}
\textrm{Im} \Sigma(k,\omega_+) &=
                           \int \frac{\textrm{d}^3 k'}{(2\pi)^3}
                           \int_{-\infty}^{\infty} \frac{\textrm{d} \omega'}{2\pi}
                           \langle \mathbf{kk}' | \textrm{Im }T(\omega+\omega'_+) | \mathbf{kk}' \rangle_A
                           A(k',\omega') \nonumber \\
                           &\times \big[ f(\omega')+b(\omega+\omega')\big] \, .
\label{eq:sigma_ladder}
\end{align}
Note the presence of a Bose-Einstein distribution, 
$b(\Omega) = \left[ e^{\beta [\Omega - 2 \mu]} - 1 \right]^{-1}$, as a consequence of the
symmetric treatment of particle-particle and hole-hole correlations. 

The effective in-medium interaction is calculated in the ladder approximation. This accounts for the
repeated scattering of particles in the medium and it is well-suited for the low-density strong-interaction
regime of interest for nuclear matter \cite{fetter}. The $T$-matrix is determined by the solution of the 
integral equation:
\begin{eqnarray}
\langle \mathbf{k k'} | T (\Omega_+) | \mathbf{p p'} \rangle_A &=&
                                \langle \mathbf{k k'} | V | \mathbf{p p'} \rangle_A \nonumber \\
                                &+& \int \frac{\textrm{d}^3 q}{(2\pi)^3} \, \int \frac{\textrm{d}^3 q'}{(2\pi)^3} \,
                                \langle \mathbf{k k'} | V | \mathbf{q q'} \rangle_A \,
                                \G^0_{II}(\mathbf{q}, \mathbf{q'},\Omega_+) \nonumber \\
                                &\times& \langle \mathbf{q q'} | T (\Omega_+) | \mathbf{p p'} \rangle_A \, ,
\label{eq:tmat_sc}
\end{eqnarray}
where the intermediate propagator accounts for the propagation of two non-interacting but dressed nucleons:
\begin{eqnarray}
\G_{II}^0(k_1,k_2,\Omega_+)&=&
                                \int_{-\infty}^{\infty} \frac{\textrm{d} \omega}{2\pi} \int_{-\infty}^{\infty} \frac{\textrm{d} \omega'}{2\pi}
                                \, A(k_1,\omega) A(k_2,\omega') \frac{1 - f(\omega) - f(\omega')}{\Omega_+-\omega-\omega'} \, .
\label{eq:gII0}
\end{eqnarray}
To reduce the dimensionality of Eq.~(\ref{eq:tmat_sc}), one usually relies in the
standard partial wave decomposition. An extra simplification is achieved by using an 
angle average of the two-body propagator with respect to the center of mass and relative 
momentum of the two colliding particles \cite{frickphd,riosphd}.

Equations (\ref{eq:sigma_dec})-(\ref{eq:gII0}) form a closed set of equations that can 
be solved self-consistently. In terms of numerics, it is advantageous to work at constant
density, and therefore we supplement the previous set of equations with the normalization
of the momentum distribution:
\begin{eqnarray}
\rho = \nu \int \frac {\textrm{d}^3 k}{(2 \pi )^3 } n(k) \, ,
\label{eq:den}
\end{eqnarray}
where $\nu$ denotes the degeneracy of the system ($\nu=4$ in the case of
symmetric nuclear matter). Once convergence is reached in the self-consistent procedure
for a given temperature and density, 
one has access to the spectral function $A(k,\omega)$, 
which, loosely speaking, describes the probability of
finding a nucleon in the medium with momentum $k$ and energy $\omega$.
At this point, one can calculate several micro- and macroscopic properties of the system. 
The momentum distribution, for instance, can be computed using Eq.~(\ref{eq:nk}). The
energy per particle is also accessible from the Galitskii-Migdal-Koltun 
sum-rule:
\begin{equation}
 \frac {E}{A}(\rho,T) = \frac {\nu}{\rho} \int \frac {\textrm{d}^3 k}{(2 \pi)^3} \int_{-\infty}^{\infty}
\frac {\textrm{d} \omega}{2 \pi} \frac {1}{2} \left ( \frac {k^2}{2m} + \omega \right ) A(k,\omega) f(\omega) , 
\end{equation}
which is valid for a Hamiltonian with only two-body interactions. 
The importance of self-consistency in the calculations stems from the fact that it immediately
leads to the conservation of both micro- and macroscopic properties \cite{baym62}. In addition,
it guarantees the fulfillment of the sum-rules for the one-body spectral function \cite{frick04b,rios06b}. 

In this paper, we shall make comparisons between the finite temperature SCGF and BHF approaches.
The latter can be formally derived from the first by performing some particular approximations. 
First, one has to assume 
that, for a given momentum, all the strength of the spectral function is accumulated in one energy,
$A(k,\omega) = \delta[\omega - \varepsilon^{BHF}(k)]$, with $\varepsilon^{BHF}(k)$ the BHF single-particle 
energy. This simplifies the calculation of the non-interacting two-body propagator of 
Eq.~(\ref{eq:gII0}), which becomes a finite-temperature Pauli blocking factor involving both particle-particle and hole-hole propagation. 
Since in the BHF approach only intermediate particle-particle states are considered, the phase space factor needs to be properly modified, 
$1 - f(\omega) - f(\omega') \to [1 - f(\omega)][1 - f(\omega')]$. 
Finally, in the BHF 
self-energy, one does not take into account the contribution of the 
Bose function in Eq.~(\ref{eq:sigma_ladder}). After the BHF equations are iterated and consistency is reached, one obtains a single-particle
spectrum and an in-medium $G$-matrix interaction, the real part of which is used to obtain the energy per particle of the system:
\begin{align}
\frac{E}{A}(\rho,T) & = \frac{\nu}{\rho} \int \frac{\textrm{d}^3 k}{(2 \pi)^3} \frac{k^2}{2m} 
f[\varepsilon^{BHF}(k)] \nonumber \\
& +
\frac{\nu}{2 \rho} \int \frac{\textrm{d}^3 k}{(2 \pi)^3} \frac{\textrm{d}^3 k'}{(2 \pi)^3} 
\langle \mathbf{k k'} | \textrm{Re} \, G (\Omega = \varepsilon^{BHF}(k) + \varepsilon^{BHF}(k')_+) | \mathbf{k k'} \rangle_A \times \nonumber \\
&\times f[ \varepsilon^{BHF}(k) ] f[ \varepsilon^{BHF}(k') ] \, .
\label{eq:enerbhf}
\end{align}
The SCGF and the BHF approaches with two-body NN interactions cannot reproduce the saturation properties
of nuclear matter at zero temperature, due to the lack of repulsive contributions, most probably those
coming from three-body forces \cite{baldo99,zuo03}. The region of interest for liquid-gas
phase transition studies is however in the low-density regime (critical densities are typically 
$\frac{1}{2}$ to $\frac{1}{3}$ of $\rho_0$) and this density regime should not be strongly modified by the presence of three-body forces. Estimates of the importance of the three-body forces in the liquid-gas regime have been performed within the BHF approximation, indicating small modifications to the critical properties \cite{baldo99,zuo03}.

To assess the model-dependence of the properties of this transition, we will compare 
the results obtained with two different NN interactions, the CDBONN \cite{cdbonn} and the Argonne V18 \cite{av18}
potentials. Although both of them reproduce the scattering phase-shifts up to about 300 MeV, 
they have very different short range cores, off-shell structure and tensor components. The many-body calculations 
depend on these details and therefore the properties of dense matter, in particular the critical
properties of the liquid-gas phase transition, will be different for the two interactions. To our
knowledge, this is the first time that two different realistic interactions are used within the SCGF approach
to study the liquid-gas coexistence and the critical properties. 
In all the calculations quoted in the following and for both the 
SCGF and the BHF approximations, partial waves up to $J=8$ have been included. The in-medium 
effective interactions have been computed with $J \le 4$ and the Born approximation has been
used for $J>4$. 

\section{Thermodynamical properties of nuclear matter}

A complete TD description of the system requires the computation of the free energy,
$F= E - T S$. As indicated in the previous section, the internal energy in the SCGF approach can be calculated 
from the one-body propagator via the Galitskii-Migdal-Koltun sum-rule. 
Therefore, a suitable method for the calculation of the entropy is required to 
describe the thermodynamics of the system. 
The LW formalism can be used to find an expression
of the grand-canonical potential, $\Omega$, in terms of dressed propagators \cite{carneiro73,carneiro75,luttinger60a}. 
Because this expression of $\Omega$ is stationary with respect to variations of the one-body Green's functions, 
one can easily compute the
entropy from the derivative $S = - \left. \frac{\partial \Omega}{\partial T} \right|_\mu$. This entropy
can be split in two terms, $S=S^{DQ}+S'$. The first one is the so-called dynamical 
quasi-particle (DQ) entropy density:
\begin{eqnarray}
S^{DQ} = \nu \int \frac {\textrm{d}^3 k}{(2 \pi )^3} \int_{-\infty}^{\infty} \frac {\textrm{d} \omega}{2 \pi} \sigma (\omega) 
B(k,\omega) \, ,  
\label{eq:quasientro}
\end{eqnarray}
defined as the convolution of a statistical factor, $\sigma(\omega) = -f(\omega) \ln f(\omega) - [1 - f(\omega)] \ln [1 -f(\omega)]$,
and a spectral function $B(k,\omega)$, related to the single-particle spectral function, $A(k,\omega)$, and the self-energy
by the following equation:
\begin{eqnarray}
B(k,\omega) = A(k,\omega) \left [ 1 - \frac {\partial \textrm{Re} \Sigma (k,\omega)}{\partial \omega} \right ] - 2 
\frac {\partial \textrm{Re} \G(k,\omega)}{\partial \omega}  \textrm{Im} \Sigma (k,\omega_+) \, .
\end{eqnarray}
This expression takes into account the correlations of the dressed particles in the medium, since it
includes finite width effects. In this paper, we shall make the assumption that $S'$ is 
negligible. As a matter of fact, it has been shown by Carneiro and Pethick that its effects are constrained
by phase space restrictions \cite{carneiro75}. This assumption is also confirmed by the fact that the
results ignoring the contribution of $S'$ are thermodynamically consistent \cite{rios06}, as we shall see in the following. The free energy will be computed from the difference of the Galitskii--Migdal--Koltun
sum-rule energy and the DQ entropy, $F=E^{GMK}-TS^{DQ}$. In the BHF approach, the entropy should not include any effect due to the widening of the quasi-particle peak and therefore it will be computed from the mean-field expression:
\begin{equation}
S^{BHF} =  \nu \int \frac {\textrm{d}^3 k}{(2 \pi )^3} \sigma\left[ \varepsilon^{BHF}(k) \right] \, .
\label{eq:entrobhf}
\end{equation}

\begin{figure}[t]
	\begin{center}
		\includegraphics[width=12cm]{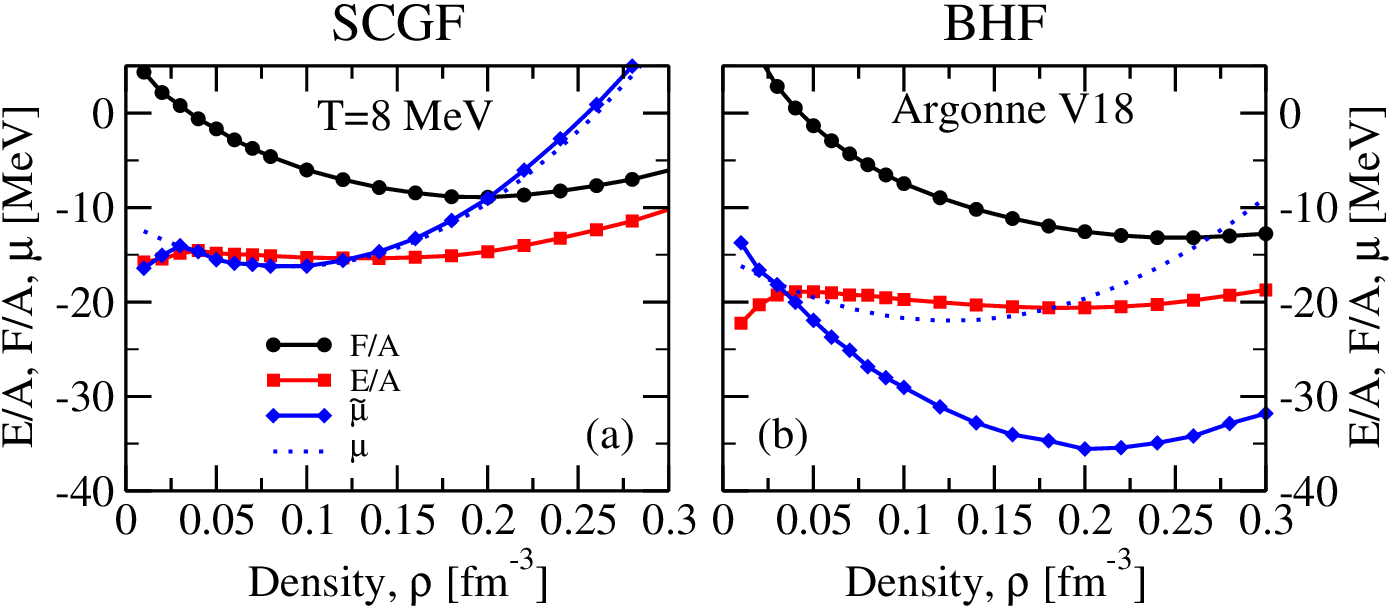}  
		\caption{(Color online) Energy per particle (circles), free energy per 
 particle (squares) and microscopic chemical potentials $\tilde \mu$ (diamonds) for the SCGF approach (panel (a)) and the BHF approach (panel (b)) as a function of the density at $T=8$ MeV for the Argonne V18 potential. The macroscopic chemical potential $\mu$ is also shown (dotted line). }
		\label{fig:TD_AV}
	\end{center}
\end{figure}

\begin{figure}[t]
	\begin{center}
		\includegraphics[width=12cm]{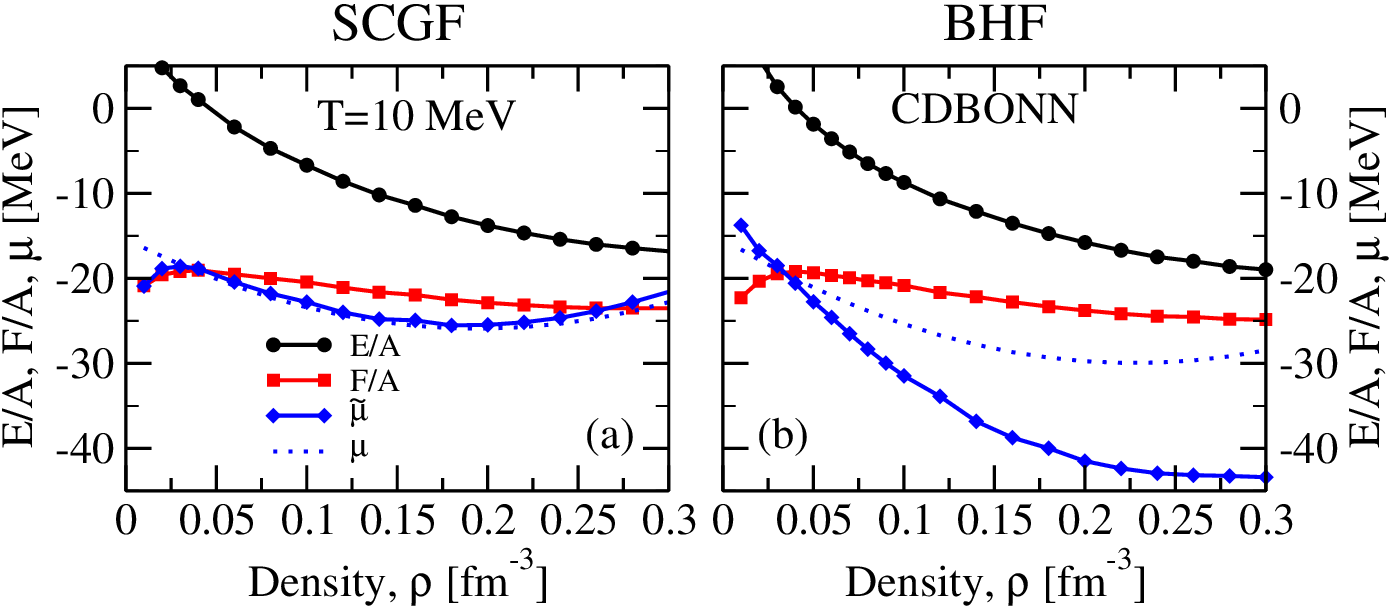}  
		\caption{(Color online) Same as Fig. 1, but for the CDBONN potential at $T=10$ MeV.}
		\label{fig:TD_CD}
	\end{center}
\end{figure}

The free energy per particle, together with the energy per 
particle and the chemical potential, are shown in Figs.~\ref{fig:TD_AV} and \ref{fig:TD_CD} as a function of the 
density at constant temperature. 
In Fig.~\ref{fig:TD_AV} we display the results at $T=8$ MeV for the Argonne V18 interaction, while in
Fig.~\ref{fig:TD_CD} we consider the CDBONN interaction at $T=10$ MeV. Both temperatures are sufficiently below 
the corresponding critical temperatures, as we will see. Note that in both figures
panel (a) corresponds to the SCGF results and panel (b) to the BHF ones. The latter have been obtained
from the combination of Eq.~(\ref{eq:enerbhf}) for the energy and Eq.~(\ref{eq:entrobhf}) for the entropy.
Let us first discuss the differences between the two many-body approaches. It is already well-established 
that hole-hole propagation, which is included in the SCGF but not in the BHF
approach, yields a repulsive contribution to the energy per particle of
about $4-6$ MeV close to saturation density \cite{dewulf03,frick03}. Since this
repulsive contribution tends to increase with nuclear density, one obtains a
smaller saturation density in the SCGF as compared to the BHF approach. 
A similar effect is also observed for the free energy 
per particle, since both the dynamical quasi-particle and the BHF entropies are quite close
to each other \cite{rios06}.
 
The temperatures considered in Figs.~\ref{fig:TD_AV} and \ref{fig:TD_CD} are slightly different
from each other, so one should be cautious when comparing these results. 
Nevertheless, this comparison exhibits the main features
which have frequently been discussed in the literature for the energy as a
function of density at zero temperature:
the CDBONN interaction contains weaker tensor components than the Argonne V18.
This implies that the density dependent suppression effects in the iterated
tensor terms are less efficient for the former interaction than for the latter.
This leads to a more attractive energy per nucleon and a larger saturation
density for the CDBONN than for Argonne V18 interaction, features which are also
present in Figs.~\ref{fig:TD_AV} and \ref{fig:TD_CD}.

It is interesting to compare the free energy and the chemical potential to check the fulfillment of TD
consistency. Some properties of the system can be 
computed either microscopically (from Green's function theory) or macroscopically (from
the TD properties of the system). A TD consistent many-body approximation will yield the same 
result for both of them. A very sensitive quantity to this test is the chemical potential.
On the one hand, it can be computed microscopically from the normalization of the 
momentum distribution, Eq.~(\ref{eq:den}), giving rise to the microscopic chemical potential,
$\tilde \mu$ (diamonds in Figs.~\ref{fig:TD_AV} and \ref{fig:TD_CD}). 
On the other hand, one can compute it from the derivative of the free energy 
with respect to the number of nucleons  at constant temperature, 
$\mu = \frac{\partial F }{\partial N }$ (dotted lines).
The differences between $\tilde \mu$ and $\mu$ for the BHF approach can be larger
than $15$ MeV, showing its lack of consistency. Note that, in particular,
the Hugenholtz-van Hove theorem is violated, \emph{i.e.} $\tilde \mu$ 
does not coincide with $\frac{F}{A}$ at its minimum. The violation seems to 
be larger for CDBONN, $\sim 20$ MeV, than for Argonne V18, $\sim 10$ MeV.
The SCGF results, however, fulfill 
TD consistency with less than one MeV error in a wide density range for both interactions. 
Note that
$\mu$ has been computed by fitting a fourth-order polynomial to the free energy density
and determining the derivatives from this polynomial fit. The lack of accuracy in the fit is responsible for the 
deviations at very low densities. In any case, the numerical implementation of the 
ladder approximation by means of the SCGF scheme leads to TD consistent results, 
independently of the NN potential under consideration \cite{bozek01}. 

\begin{figure}[t]
	\begin{center}
		\includegraphics[width=12cm]{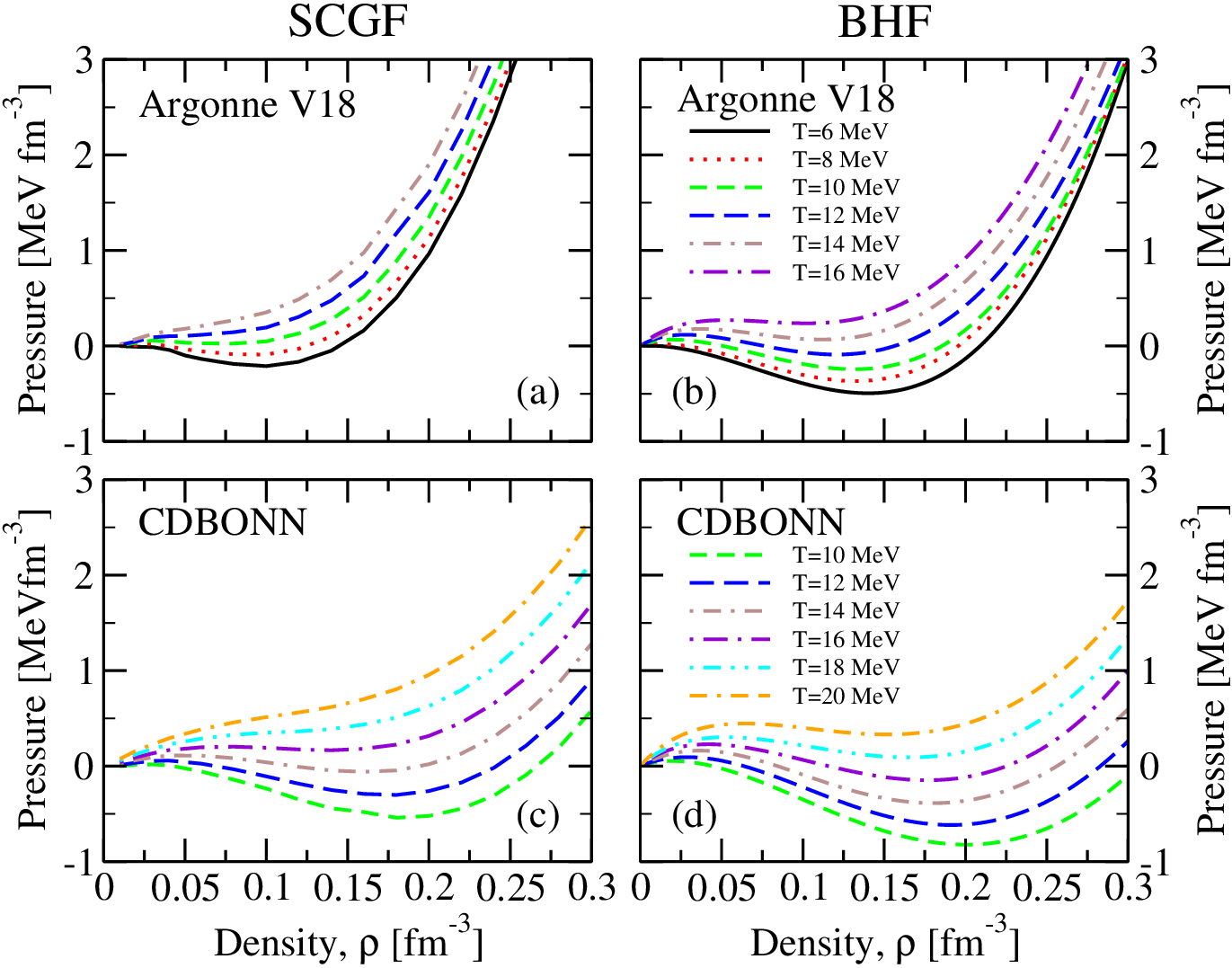}  
		\caption{(Color online) Pressure as a function of density for several temperatures, obtained from the SCGF and BHF calculations with the Argonne V18 and CDBONN interactions.}
		\label{fig:pres_cd_av18}
	\end{center}
\end{figure}

The pressure is shown as a function of the density in Fig. \ref{fig:pres_cd_av18}
for several temperatures and for both the 
SCGF (left panels)  and  BHF (right panels) approaches. 
The upper panels, (a) and (b), correspond to Argonne V18
and the lower ones, (c) and (d), to CDBONN. The pressure is obtained from the TD relation,
$p= \rho (\mu - F/A)$. Due to the conserving properties of the SCGF approach, 
we can compute the pressure at every density and temperature by using the microscopic 
chemical potential in the previous
expression. For BHF, however, $\mu$ has to be computed as a numerical 
derivative of the free energy with
respect to $\rho$. Note that numerical problems can appear in the low density limit, 
due to logarithmic density dependences in this region.  

In general, one can say that the repulsive effect of the hole-hole propagation
in the free-energy, which we have already discussed above, 
is translated into larger pressures in the SCGF  
as compared to the BHF approach, especially at large densities. 
The SCGF method, therefore, yields a stiffer equation of state than the BHF approach. 
By construction, the TD chemical potential, $\mu$, crosses the free-energy curve at its
minimum, thus yielding a point of zero pressure. This defines the saturation density at each temperature.
The repulsive effect induced by the propagation of holes in the SCGF approach is reflected in 
a smaller saturation density with respect to BHF, \emph{i.e.} the minimum of the free energy is 
shifted to smaller densities in the SCGF approach. This effect has already been observed
at zero temperature \cite{dewulf03} and it appears to hold when thermal effects are
taken into account. Above a certain temperature, the minimum of the free energy 
per particle disappears and the equation $P(\rho)=0$ has no solution anymore. 
This defines the so-called flashing temperature, $T_{f}$.
The first column of Table~\ref{tab:critical} gives the flashing temperature for the different
approaches and potentials. The SCGF results lead to $T_f$'s which are about
$3$ MeV lower than the BHF ones. 

The differences in pressures from the two many-body approaches are sizeable, but the
differences due to the change of potentials are even larger.
The pressure for Argonne V18, for instance, increases much more steeply with density than
CDBONN does. There are also substantial differences in the saturation densities induced
by the two potentials at all temperatures, with Argonne V18 leading to lower saturation
densities than CDBONN. This is in agreement with the fact that at $T=0$ the BHF saturation density
is much larger for CDBONN than for Argonne V18 \cite{muther00}. Note also that the temperature
ranges explored in the upper and lower panels of Fig.~\ref{fig:pres_cd_av18} are not the same.
The temperature dependences induced by the two potentials are therefore
rather different. For instance, the flashing temperatures for Argonne V18 are about $4-5$ 
MeV lower than for CDBONN with both SCGF and BHF. This suggests that the different off-shell and tensor
components of the NN forces do not only affect the properties of nuclear matter 
at zero temperature, but also its TD properties in an important way. 
It remains to be seen if the experimental knowledge gathered about the thermodynamical
properties of nuclear systems can provide additional information to constrain the NN interaction
and select the proper many-body approach to be used in their description. 

For both approaches and potentials, the pressure decreases with density
in a given range. This signals the existence of a mechanical instability, which
is associated to a first order liquid-gas phase transition. The properties of this
transition are studied in the following. 

\section{Liquid-gas phase transition}

\begin{figure}[t]%
  \begin{center}
    \includegraphics[width=12cm]{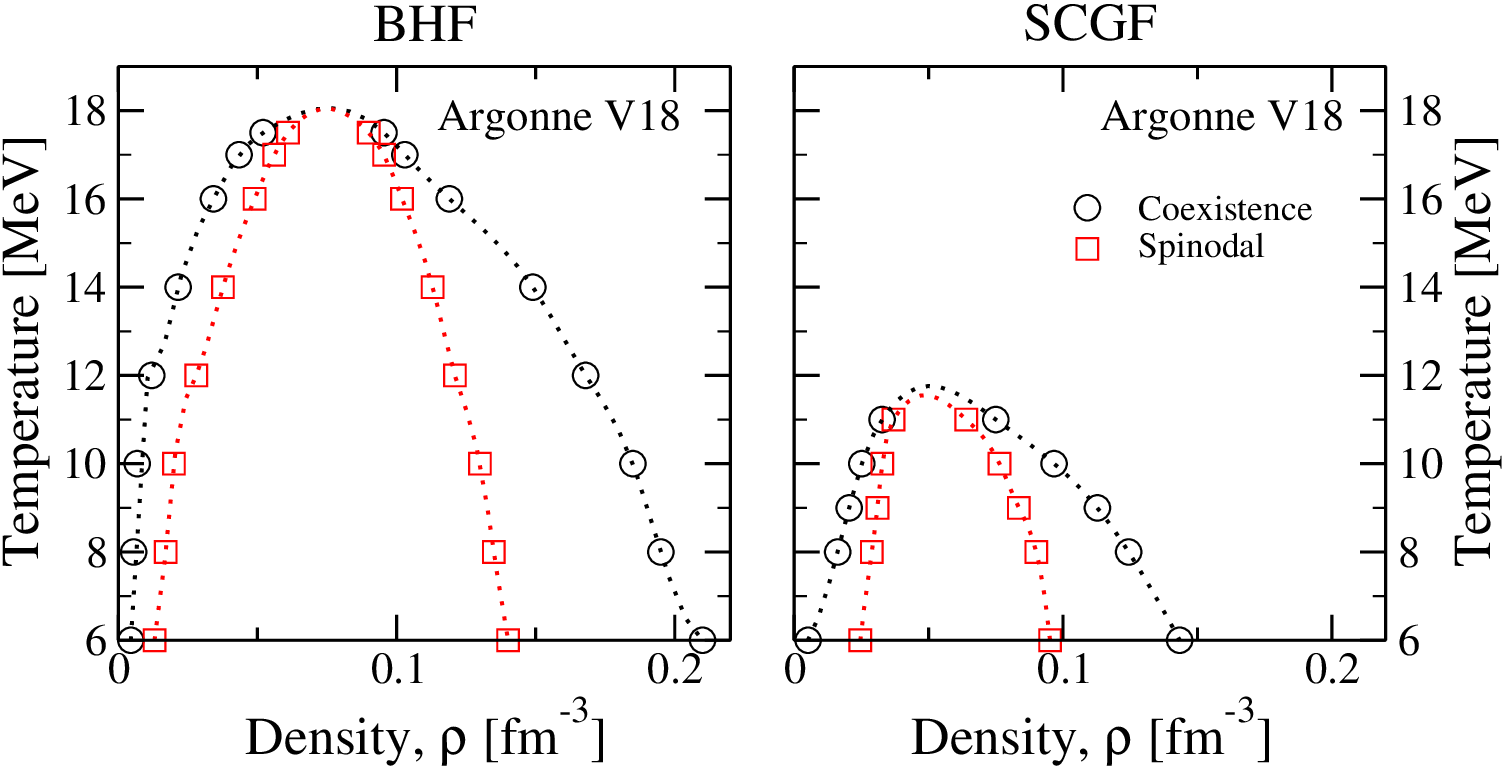}
    \caption{(Color online) Coexistence (circles) and spinodal (squares) lines for symmetric nuclear matter 
      within the BHF (left panel) and SCGF (right panel) approaches for the Argonne V18 interaction.}
    \label{fig:coex_av18}
  \end{center}
\end{figure}

\begin{figure}[t]%
  \begin{center}
    \includegraphics[width=12cm]{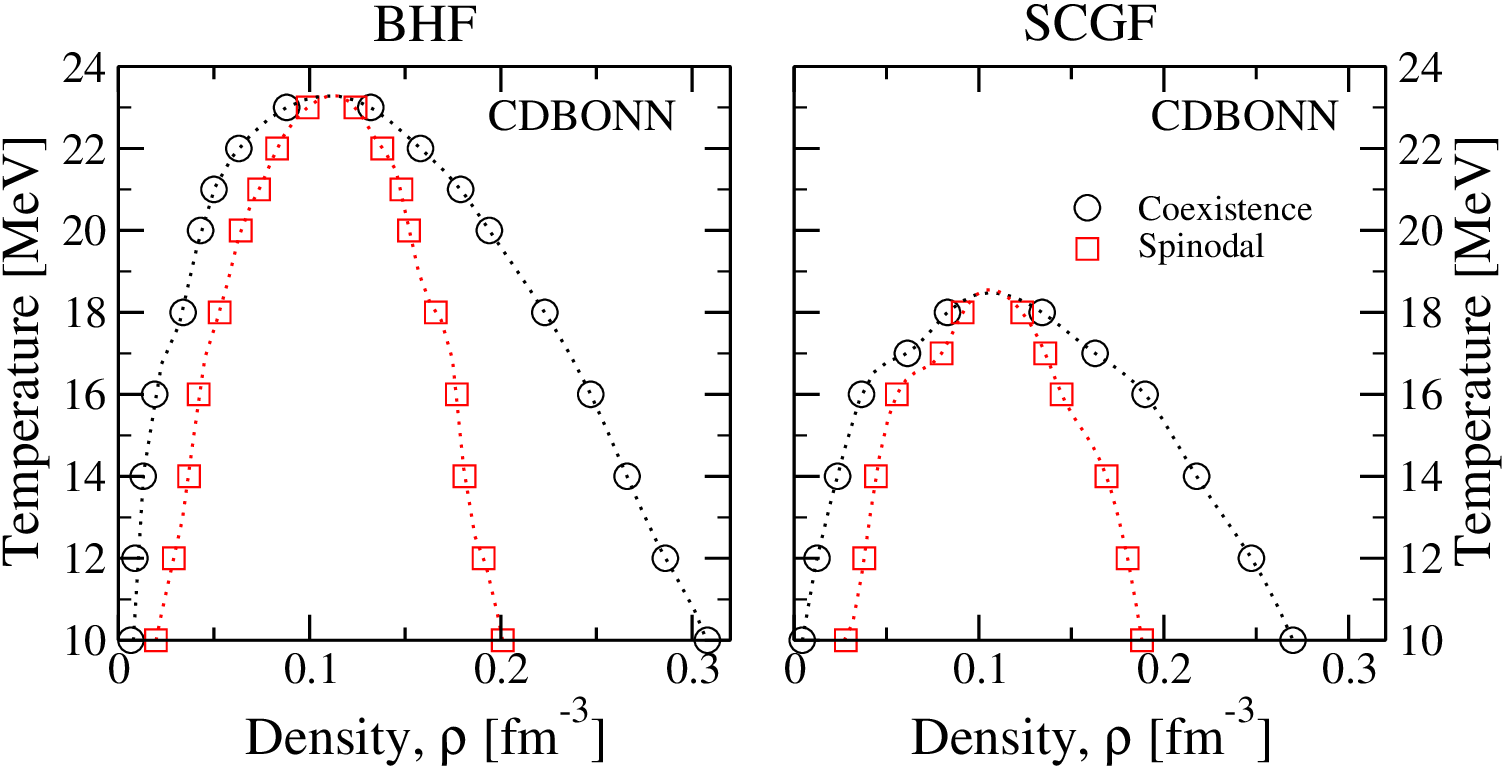}
    \caption{(Color online) Same as Fig. 4, but for the CDBONN interaction.}
    \label{fig:coex_cdbonn}
  \end{center}
\end{figure}

A physical interpretation of the TD unstable zone is customarily obtained by making use of the
Maxwell construction. For each temperature, one should find the gas and liquid density, for which
the equations $\mu(\rho_g) = \mu(\rho_l)$ and $p(\rho_g)=p(\rho_l)$ are simultaneously 
satisfied. For a given temperature, the range $\rho_g-\rho_l$ gives the coexistence region, 
where the gas and liquid phases coexist at constant pressure and chemical potentials.
The spinodal region is defined by the violation of the TD stability criteria, 
$\frac{\partial \mu}{\partial \rho} > 0$ and $\frac{\partial p}{\partial \rho} > 0$. 
For a one component system, both conditions are equivalent. This spinodal region 
lies within the liquid-gas coexistence region in the density-temperature plane, and  
the region between the two curves defines the so called metastable region.
Finding the spinodal and coexistence densities at each temperature, one obtains the phase diagrams   
shown in Fig.~\ref{fig:coex_av18} for Argonne V18 and in Fig.~\ref{fig:coex_cdbonn} for CDBONN. 
Note that the two figures have different vertical scales, because of the
large differences in the TD properties induced by the two interactions. 
Once again, let us emphasize that one should be cautious when using fitting 
procedures for the free energy. These are necessary to obtain suitable analytical expressions
for the derivatives of $F(\rho,T)$, needed to implement the 
liquid-gas coexistence conditions. It is also worth noting that,
for the temperatures considered, both methods (BHF and SCGF) are numerically stable 
in a range of densities which covers from very low ones (gas phase, $\rho=0.01$ fm$^{-3}$) to 
relatively large ones (liquid phase, $\rho=0.30$ fm$^{-3}$). 
For the SCGF approach, numerical difficulties related to the pairing
instability appear at lower temperatures \cite{vonder93,bozek99}.

The critical temperature for the liquid-gas phase transition of symmetric nuclear matter corresponds 
to the maximum of the spinodal and coexistence lines, which coincide with each other at the critical
point. The flashing temperature, $T_f$, always lies below $T_c$ and, as commented above, correspond to the maximum temperature at which the pressure presents a node. $T_f$ represents the highest temperature that finite nuclei can withstand without thermally dissociating.
The critical properties as well as the limiting temperatures for the two approaches 
and interactions
are listed in Table~\ref{tab:critical}. Both $T_f$ and $T_c$ depend strongly on the type
of approximation used to study nuclear matter, and also on the NN interaction. 
Note that, for a given potential, BHF results always lead to larger flashing and critical 
temperatures (about $3-4$ MeV larger) than SCGF. Also, the results
for Argonne V18 are, within each approximation, about $4-5$ MeV lower than those of 
CDBONN.

At this point we are going to compare our predictions for the critical
temperature with those obtained in other approximation schemes.
We obtain the largest 
critical temperature for the CDBONN potential in the BHF approximation, with a 
value of $T_c = 23.3$ MeV. 
The critical temperature for CDBONN using the SCGF 
approach is, however, $T_c=18.5$ MeV. This is in close agreement with the value of 
$T_c=18$ MeV obtained with the somewhat similar Bonn B potential in the Bloch-de Dominicis formalism \cite{baldo04}.
The Argonne V18 interaction yields lower critical temperatures when compared to CDBONN, with $T_c=18.1$
MeV for BHF and $T_c=11.6$ MeV for SCGF. The first result can be compared with calculations 
performed with other approximations and the same NN potential. The LOCV results of Ref.~\cite{moshfegh05} 
correspond to a 
somewhat larger critical temperature of $T_c=22.2$ MeV for V18, while a more recent
variational calculation with frozen correlations and three-body forces leads to $T_c=18$ MeV 
\cite{kanzawa07}, rather close to our BHF result. 
The authors of Ref.~\cite{zuo03} found a critical temperature
$T_c=16$ MeV when the effect of three-body forces was neglected.
Other results in the range $16-20$ MeV have been obtained with similar interactions. 
One can estimate the critical temperature of the variational calculation by Friedman and 
Pandharipande with the Urbana
V14 potential (plus three-body force) to be about $T_c=17-18$ MeV \cite{friedman81}.
The Bloch-de Dominicis calculation of Ref.~\cite{baldo99} for the Argonne V14 interaction
yields a critical temperature of
$T_c=21$ MeV with two-body forces and $T_c=20$ MeV when the effect of three-body forces is included.
In all these cases, the critical temperature is about $7-9$  MeV larger than the one obtained with Argonne V18 in the SCGF
approach. This is in fact below the usually quoted critical temperature for infinite matter, of around 
$15-20$ MeV. Yet, some other models, especially the relativistic ones, have found similar low values of $T_c$.
For the Density Dependent Relativistic
Mean-Field calculation of Ref.~\cite{hua00}, critical temperatures of the order of $12$ MeV were found. The 
Dirac--Brueckner--Hartree--Fock calculations yield
also low critical temperatures, like the $T_c=12$ MeV and $T_c=10.4$ MeV of Refs.~\cite{terhaar86} and \cite{huber98},
respectively, obtained using One-Boson-Exchange interactions fitted to NN data.  
The non-relativistic semirealistic model of Ref.~\cite{das92} found convergence problems, possibly associated to the
liquid-gas transition, below a critical temperature as low as $9$ MeV. 

Can one explain in simple terms the origin of the large differences in critical temperatures between the
two approaches and NN interactions? In fact, there are some simple models which try to relate the critical
properties to the saturation properties of nuclear matter at zero temperature 
\cite{jaqaman83,suraud}. A particularly useful and simple estimate is obtained from the Kapusta 
model \cite{kapusta84}, which supposes that the temperature dependences are quadratic 
(as in the free Fermi gas close to the degenerate limit) and modulated by an effective mass,
$m^*$, which governs the density of states. Under the additional assumption that the
zero temperature energy per particle can be characterized by the compressibility, $K$, one finds that:
\begin{equation}
T_c = 0.326 \left( \frac{K}{m^*} \right)^{1/2} \rho_0^{1/3} \, ,
\label{eq:kapusta}
\end{equation}
\emph{i.e.} the critical temperature increases with
$K$ and the saturation density, $\rho_0$. The presence of the pairing instability
\cite{vonder93,bozek99} prevents us from decreasing the temperature in the SCGF 
scheme below about $5$ MeV and we cannot safely extrapolate the values 
of $K$, $m^*$ and $\rho_0$ to the zero temperature limit.
We have, however,
performed BHF calculations at zero temperature and found the compressibility
$K=279$ MeV ($K=212$ MeV) and the saturation density $\rho_0=0.35$ fm$^{-3}$ ($\rho=0.23$ fm$^{-3}$)
for CDBONN (Argonne V18). The estimates obtained with Eq.~(\ref{eq:kapusta}) with a free
mass lead to the critical temperatures $T_c \sim 25$ MeV and $T_c \sim 19$ MeV,
in rather good agreement with the values reported in Table \ref{tab:critical}. 
The available results of SCGF calculations seem to indicate that the saturation point 
decreases with respect to the BHF calculations; the compressibility,
however, increases \cite{dewulf03}. Within the Kapusta model these two features
would essentially compensate each other and predict a critical temperature for
the SCGF approach which is about the same as the one obtained in the BHF
approximation. Note, however, that the SCGF approach leads to an enhancement of
the density of states at low excitation energies which could be described in
terms of a larger effective mass $m^*$ as compared to BHF~\cite{frick02}. 

Independently of this analysis, it is generally true that the critical temperature is 
correlated and somewhat close to the binding energy at saturation, 
$T_c \sim E/A$ \cite{jaqaman83}.
The binding energies with BHF at zero temperatures are $E/A=21.6$ MeV for CDBONN and 
$E/A=16.2$ MeV for Argonne V18. Note that the $4-5$ MeV difference in those values is close to the difference 
observed in the critical temperatures. Now, since usually the SCGF leads to binding energies which are $4-6$ MeV 
more repulsive than those of the BHF approach, this simple argument would suggest that the critical 
temperatures should also decrease by a similar amount, as observed in Table \ref{tab:critical}.

The Kapusta model also predicts the value of the critical density, $\frac{\rho_c}{\rho_0}=\frac{5}{12}=0.417$. 
For the BHF calculations, we find $\frac{\rho_c}{\rho_0}=0.35$ for Argonne V18 and $\frac{\rho_c}{\rho_0}=0.31$ for CDBONN,
which are somewhat closer to the empirical formula  $\frac{\rho_c}{\rho_0}=\frac{1}{3}$ \cite{jaqaman83}.
Note also that the value for Argonne V18 in the BHF approximation lies within the range $0.07-0.09$ fm$^{-3}$ 
quoted in Ref.~\cite{baldo99} for the V14 potential. 
The spread in critical densities is significant when we change from one potential to another,
but it does not differ so much when we consider different approximations. 
The differences
are however more drastic for the critical pressure, which changes by almost an order of magnitude when
comparing different approximations and potentials. Yet, surprisingly, the discrepancies are somewhat 
less important for the
adimensional parameter $\frac{p_c}{\rho_c T_c}$. For a van der Waals equation of state, this parameter is
$\frac{3}{8}=0.375$. The results of the last column of Table $\ref{tab:critical}$ are quite 
below this value. Intriguingly, the BHF results seem to lead to the same value, in spite of
the large differences in each of their critical parameters.

Finally, let us recall again that we have not considered any three-body forces in the previous calculations.
In terms of the phase diagram, one expects
that the inclusion of three-body forces will shift the liquid 
coexistence branch to lower densities in the low temperature phase, 
but will presumably have a small effect at large temperatures. 
The effects on the gas phase, if any, would probably be very small in the homogeneous case.
In fact, previous evaluations of the three-body effects on the critical temperature
obtained within the BHF approach 
seem to indicate that these are rather small, about $1-3$ MeV \cite{baldo99,zuo03}.
Such a decrease is much smaller than the discrepancies observed here when changing 
the two-body interaction or the many-body method.

\section{Summary and Conclusions}

We have calculated the TD properties of symmetric nuclear matter within the Self-Consistent
Green's Function and the Brueckner--Hartree--Fock approaches for two realistic NN interactions,
the CDBONN and the Argonne V18 potentials. The calculations cover a wide range of densities and
temperatures. In the SCGF-LW approach, the entropy has been computed within the 
dynamical quasi-particle approximation, which takes into account the effects of 
correlations in the width of the quasi-particle peak.
A very good agreement between the microscopic and macroscopic chemical potentials 
is found, highlighting the TD consistency of the SCGF-LW approach at
the numerical level. This is in contrast to the BHF approximation,
which yields a violation of the Hugenholtz-van Hove theorem by $10$ MeV ($20$ MeV) 
for the Argonne (CDBONN) potential.

The essential difference between SCGF and BHF is the consistent 
inclusion of hole-hole propagation terms in the former approach, leading to
non-trivial spectral distribution functions and partial occupation probabilities
for states with momenta above and below the Fermi momentum. These hole-hole
terms tend to provide some repulsion, which increases with density. This feature
is also reflected in the calculated pressure, for which the SCGF approach yields
larger values, in particular at higher densities. This implies that the
equation of state derived within the SCGF approach tends to be stiffer than the
corresponding one evaluated within the BHF approximation. The repulsive effect
of the hole-hole terms also leads to a lower flashing temperatures for the SCGF
approach as compared to BHF.

When comparing the results between two different NN interactions, we
also find substantial differences, larger than those induced by the use of different
many-body approaches. In particular, the Argonne V18 interaction leads to a 
stiffer equation of state and a lower flashing temperature than CDBONN.
 
The liquid-gas phase diagram for nuclear matter has been studied for the first time in the
framework of the microscopic SCGF approach for two realistic NN interactions and
critically compared with the results obtained in the BHF approach. Substantial differences for the critical
properties are found when changing the potential and the many-body approximation. 
The SCGF leads to critical temperatures which are $5-7$ MeV lower than those obtained
with the BHF approximation. Within the same approximation, CDBONN leads to results
which are $6-7$ MeV larger. For BHF, where $T=0$ calculations can be performed safely,
we find that the critical density is about a third of saturation density and that the
critical temperature can be well approximated by Eq.~(\ref{eq:kapusta}), in terms of 
the compressibility and the saturation density. 

Of course, to have a proper estimation of the critical temperature for finite nuclei, one 
should also take into account Coulomb effects and the existence of a surface tension.
These results would further reduce the critical temperature, by a factor of $1/2-1/4$ 
\cite{natowitz02,baldo04}. 
In this paper we have found that realistic calculations allow for a large range
of critical temperatures, in the same way that they predict different saturation 
properties.
In particular, the important reduction in critical temperatures found for
the Argonne interaction might be relevant when trying to connect
the data of multifragmentation reactions with the liquid-gas phase transition for bulk 
nuclear matter. 

\section{Acknowledgments}
The authors are grateful to Isaac Vida\~na for useful and stimulating discussions
and for the use of his BHF codes.
This work was partially supported by the NSF under Grant No. PHY-0555893
and by the MEC (Spain) and FEDER under Grant No. FIS2005-03142 and 
2005SGR-00343 (Generalitat de Catalunya). 

\bibliographystyle{apsrev}
\bibliography{main}

\begin{thebibliography}{63}
\expandafter\ifx\csname natexlab\endcsname\relax\def\natexlab#1{#1}\fi
\expandafter\ifx\csname bibnamefont\endcsname\relax
  \def\bibnamefont#1{#1}\fi
\expandafter\ifx\csname bibfnamefont\endcsname\relax
  \def\bibfnamefont#1{#1}\fi
\expandafter\ifx\csname citenamefont\endcsname\relax
  \def\citenamefont#1{#1}\fi
\expandafter\ifx\csname url\endcsname\relax
  \def\url#1{\texttt{#1}}\fi
\expandafter\ifx\csname urlprefix\endcsname\relax\def\urlprefix{URL }\fi
\providecommand{\bibinfo}[2]{#2}
\providecommand{\eprint}[2][]{\url{#2}}

\bibitem[{\citenamefont{Prakash et~al.}(1996)\citenamefont{Prakash, Bombaci,
  Prakash, Ellis, Lattimer, and Knorren}}]{prakash96}
\bibinfo{author}{\bibfnamefont{M.}~\bibnamefont{Prakash}},
  \bibinfo{author}{\bibfnamefont{I.}~\bibnamefont{Bombaci}},
  \bibinfo{author}{\bibfnamefont{M.}~\bibnamefont{Prakash}},
  \bibinfo{author}{\bibfnamefont{P.~J.} \bibnamefont{Ellis}},
  \bibinfo{author}{\bibfnamefont{J.~M.} \bibnamefont{Lattimer}},
  \bibnamefont{and} \bibinfo{author}{\bibfnamefont{R.}~\bibnamefont{Knorren}},
  \bibinfo{journal}{Phys. Rep.} \textbf{\bibinfo{volume}{280}},
  \bibinfo{pages}{1} (\bibinfo{year}{1996}).

\bibitem[{\citenamefont{Oechslin and Janka}(2007)}]{janka07}
\bibinfo{author}{\bibfnamefont{R.}~\bibnamefont{Oechslin}} \bibnamefont{and}
  \bibinfo{author}{\bibfnamefont{H.-T.} \bibnamefont{Janka}},
  \bibinfo{journal}{Phys. Rev. Lett.} \textbf{\bibinfo{volume}{99}},
  \bibinfo{pages}{121102} (\bibinfo{year}{2007}).

\bibitem[{\citenamefont{Yakovlev and Pethick}(2004)}]{yakovlev04}
\bibinfo{author}{\bibfnamefont{D.~G.} \bibnamefont{Yakovlev}} \bibnamefont{and}
  \bibinfo{author}{\bibfnamefont{C.~J.} \bibnamefont{Pethick}},
  \bibinfo{journal}{Annu. Rev. Astron. Astrophys.}
  \textbf{\bibinfo{volume}{42}}, \bibinfo{pages}{169} (\bibinfo{year}{2004}).

\bibitem[{\citenamefont{Pochodzalla}(1997)}]{pochodzalla97}
\bibinfo{author}{\bibfnamefont{J.}~\bibnamefont{Pochodzalla}},
  \bibinfo{journal}{Prog. Part. Nucl. Phys.} \textbf{\bibinfo{volume}{39}},
  \bibinfo{pages}{443} (\bibinfo{year}{1997}).

\bibitem[{\citenamefont{Siemens}(1983)}]{siemens83}
\bibinfo{author}{\bibfnamefont{P.~J.} \bibnamefont{Siemens}},
  \bibinfo{journal}{Nature} \textbf{\bibinfo{volume}{305}},
  \bibinfo{pages}{410} (\bibinfo{year}{1983}).

\bibitem[{\citenamefont{Danielewicz et~al.}(2002)}]{dani02}
\bibinfo{author}{\bibfnamefont{P.}~\bibnamefont{Danielewicz}}
  \bibnamefont{et~al.}, \bibinfo{journal}{Science}
  \textbf{\bibinfo{volume}{298}}, \bibinfo{pages}{1592} (\bibinfo{year}{2002}).

\bibitem[{\citenamefont{Das et~al.}(2005)}]{das05}
\bibinfo{author}{\bibfnamefont{C.~B.} \bibnamefont{Das}} \bibnamefont{et~al.},
  \bibinfo{journal}{Phys. Rep.} \textbf{\bibinfo{volume}{406}},
  \bibinfo{pages}{1} (\bibinfo{year}{2005}).

\bibitem[{\citenamefont{Pochodzalla et~al.}(1995)}]{pochodzalla95}
\bibinfo{author}{\bibfnamefont{J.}~\bibnamefont{Pochodzalla}}
  \bibnamefont{et~al.}, \bibinfo{journal}{Phys. Rev. Lett.}
  \textbf{\bibinfo{volume}{75}}, \bibinfo{pages}{1040} (\bibinfo{year}{1995}).

\bibitem[{\citenamefont{Natowitz et~al.}(2002)}]{natowitz02}
\bibinfo{author}{\bibfnamefont{J.~B.} \bibnamefont{Natowitz}}
  \bibnamefont{et~al.}, \bibinfo{journal}{Phys. Rev. C}
  \textbf{\bibinfo{volume}{65}}, \bibinfo{pages}{034618}
  (\bibinfo{year}{2002}).

\bibitem[{\citenamefont{Sobotka et~al.}(2004)\citenamefont{Sobotka, Charity,
  Toke, and Schr{\"o}der}}]{sobotka04}
\bibinfo{author}{\bibfnamefont{L.~G.} \bibnamefont{Sobotka}},
  \bibinfo{author}{\bibfnamefont{R.~J.} \bibnamefont{Charity}},
  \bibinfo{author}{\bibfnamefont{J.}~\bibnamefont{Toke}}, \bibnamefont{and}
  \bibinfo{author}{\bibfnamefont{W.~U.} \bibnamefont{Schr{\"o}der}},
  \bibinfo{journal}{Phys. Rev. Lett.} \textbf{\bibinfo{volume}{93}},
  \bibinfo{pages}{132702} (\bibinfo{year}{2004}).

\bibitem[{\citenamefont{Gupta et~al.}(2006)\citenamefont{Gupta, Mekjian, and
  Tsang}}]{tsang06}
\bibinfo{author}{\bibfnamefont{S.~D.} \bibnamefont{Gupta}},
  \bibinfo{author}{\bibfnamefont{A.~Z.} \bibnamefont{Mekjian}},
  \bibnamefont{and} \bibinfo{author}{\bibfnamefont{M.~B.} \bibnamefont{Tsang}},
  \emph{\bibinfo{title}{Advances in Nuclear Physics 26}}
  (\bibinfo{publisher}{Springer US}, \bibinfo{year}{2006}), chap.
  \bibinfo{chapter}{Liquid-gas phase transition in nuclear multifragmentation},
  p.~\bibinfo{pages}{89}.

\bibitem[{\citenamefont{Levit and Bonche}(1985)}]{levit85}
\bibinfo{author}{\bibfnamefont{S.}~\bibnamefont{Levit}} \bibnamefont{and}
  \bibinfo{author}{\bibfnamefont{P.}~\bibnamefont{Bonche}},
  \bibinfo{journal}{Nucl. Phys. A} \textbf{\bibinfo{volume}{437}},
  \bibinfo{pages}{426} (\bibinfo{year}{1985}).

\bibitem[{\citenamefont{Baldo et~al.}(2004)\citenamefont{Baldo, Ferreira, and
  Nicotra}}]{baldo04}
\bibinfo{author}{\bibfnamefont{M.}~\bibnamefont{Baldo}},
  \bibinfo{author}{\bibfnamefont{L.~S.} \bibnamefont{Ferreira}},
  \bibnamefont{and} \bibinfo{author}{\bibfnamefont{O.~E.}
  \bibnamefont{Nicotra}}, \bibinfo{journal}{Phys. Rev. C}
  \textbf{\bibinfo{volume}{69}}, \bibinfo{pages}{034321}
  (\bibinfo{year}{2004}).

\bibitem[{\citenamefont{Jaqaman et~al.}(1983)\citenamefont{Jaqaman, Mekjian,
  and Zamick}}]{jaqaman83}
\bibinfo{author}{\bibfnamefont{H.}~\bibnamefont{Jaqaman}},
  \bibinfo{author}{\bibfnamefont{A.~Z.} \bibnamefont{Mekjian}},
  \bibnamefont{and} \bibinfo{author}{\bibfnamefont{L.}~\bibnamefont{Zamick}},
  \bibinfo{journal}{Phys. Rev. C} \textbf{\bibinfo{volume}{27}},
  \bibinfo{pages}{2782} (\bibinfo{year}{1983}).

\bibitem[{\citenamefont{Kapusta}(1984)}]{kapusta84}
\bibinfo{author}{\bibfnamefont{J.}~\bibnamefont{Kapusta}},
  \bibinfo{journal}{Phys. Rev. C} \textbf{\bibinfo{volume}{29}},
  \bibinfo{pages}{1735} (\bibinfo{year}{1984}).

\bibitem[{\citenamefont{Goodman et~al.}(1984)\citenamefont{Goodman, Kapusta,
  and Mekjian}}]{goodman84}
\bibinfo{author}{\bibfnamefont{A.~L.} \bibnamefont{Goodman}},
  \bibinfo{author}{\bibfnamefont{J.~I.} \bibnamefont{Kapusta}},
  \bibnamefont{and} \bibinfo{author}{\bibfnamefont{A.~Z.}
  \bibnamefont{Mekjian}}, \bibinfo{journal}{Phys. Rev. C}
  \textbf{\bibinfo{volume}{30}}, \bibinfo{pages}{851} (\bibinfo{year}{1984}).

\bibitem[{\citenamefont{M{\"u}ller and Serot}(1995)}]{muller95}
\bibinfo{author}{\bibfnamefont{H.}~\bibnamefont{M{\"u}ller}} \bibnamefont{and}
  \bibinfo{author}{\bibfnamefont{B.~D.} \bibnamefont{Serot}},
  \bibinfo{journal}{Phys. Rev. C} \textbf{\bibinfo{volume}{52}},
  \bibinfo{pages}{2072} (\bibinfo{year}{1995}).

\bibitem[{\citenamefont{Hua et~al.}(2000)\citenamefont{Hua, Bo, and
  diToro}}]{hua00}
\bibinfo{author}{\bibfnamefont{G.}~\bibnamefont{Hua}},
  \bibinfo{author}{\bibfnamefont{L.}~\bibnamefont{Bo}}, \bibnamefont{and}
  \bibinfo{author}{\bibfnamefont{M.}~\bibnamefont{diToro}},
  \bibinfo{journal}{Phys. Rev. C} \textbf{\bibinfo{volume}{62}},
  \bibinfo{pages}{035203} (\bibinfo{year}{2000}).

\bibitem[{\citenamefont{Sil et~al.}(2004)\citenamefont{Sil, Samaddar, De, and
  Shlomo}}]{sil04}
\bibinfo{author}{\bibfnamefont{T.}~\bibnamefont{Sil}},
  \bibinfo{author}{\bibfnamefont{S.~K.} \bibnamefont{Samaddar}},
  \bibinfo{author}{\bibfnamefont{J.~N.} \bibnamefont{De}}, \bibnamefont{and}
  \bibinfo{author}{\bibfnamefont{S.}~\bibnamefont{Shlomo}},
  \bibinfo{journal}{Phys. Rev. C} \textbf{\bibinfo{volume}{69}},
  \bibinfo{pages}{014602} (\bibinfo{year}{2004}).

\bibitem[{\citenamefont{Chomaz et~al.}(2004)\citenamefont{Chomaz, Colonna, and
  Randrup}}]{cho03}
\bibinfo{author}{\bibfnamefont{P.}~\bibnamefont{Chomaz}},
  \bibinfo{author}{\bibfnamefont{M.}~\bibnamefont{Colonna}}, \bibnamefont{and}
  \bibinfo{author}{\bibfnamefont{J.}~\bibnamefont{Randrup}},
  \bibinfo{journal}{Phys. Rep.} \textbf{\bibinfo{volume}{389}},
  \bibinfo{pages}{263} (\bibinfo{year}{2004}).

\bibitem[{\citenamefont{Lejeune et~al.}(1986)\citenamefont{Lejeune, Grang\'e,
  Martzolff, and Cugnon}}]{lejeune86}
\bibinfo{author}{\bibfnamefont{A.}~\bibnamefont{Lejeune}},
  \bibinfo{author}{\bibfnamefont{P.}~\bibnamefont{Grang\'e}},
  \bibinfo{author}{\bibfnamefont{M.}~\bibnamefont{Martzolff}},
  \bibnamefont{and} \bibinfo{author}{\bibfnamefont{J.}~\bibnamefont{Cugnon}},
  \bibinfo{journal}{Nucl. Phys. A} \textbf{\bibinfo{volume}{453}},
  \bibinfo{pages}{189} (\bibinfo{year}{1986}).

\bibitem[{\citenamefont{Baldo and Ferreira}(1999)}]{baldo99}
\bibinfo{author}{\bibfnamefont{M.}~\bibnamefont{Baldo}} \bibnamefont{and}
  \bibinfo{author}{\bibfnamefont{L.~S.} \bibnamefont{Ferreira}},
  \bibinfo{journal}{Phys. Rev. C} \textbf{\bibinfo{volume}{59}},
  \bibinfo{pages}{682} (\bibinfo{year}{1999}).

\bibitem[{\citenamefont{Rios et~al.}(2005)\citenamefont{Rios, Polls, Ramos, and
  Vida{\~n}a}}]{rios05}
\bibinfo{author}{\bibfnamefont{A.}~\bibnamefont{Rios}},
  \bibinfo{author}{\bibfnamefont{A.}~\bibnamefont{Polls}},
  \bibinfo{author}{\bibfnamefont{A.}~\bibnamefont{Ramos}}, \bibnamefont{and}
  \bibinfo{author}{\bibfnamefont{I.}~\bibnamefont{Vida{\~n}a}},
  \bibinfo{journal}{Phys. Rev. C} \textbf{\bibinfo{volume}{72}},
  \bibinfo{pages}{024316} (\bibinfo{year}{2005}).

\bibitem[{\citenamefont{Friedman and Pandharipande}(1981)}]{friedman81}
\bibinfo{author}{\bibfnamefont{B.}~\bibnamefont{Friedman}} \bibnamefont{and}
  \bibinfo{author}{\bibfnamefont{V.}~\bibnamefont{Pandharipande}},
  \bibinfo{journal}{Nucl. Phys. A} \textbf{\bibinfo{volume}{361}},
  \bibinfo{pages}{502} (\bibinfo{year}{1981}).

\bibitem[{\citenamefont{Baldo et~al.}(1988)\citenamefont{Baldo, Bombaci,
  Ferreira, Giansiracusa, and Lombardo}}]{baldo88}
\bibinfo{author}{\bibfnamefont{M.}~\bibnamefont{Baldo}},
  \bibinfo{author}{\bibfnamefont{I.}~\bibnamefont{Bombaci}},
  \bibinfo{author}{\bibfnamefont{L.~S.} \bibnamefont{Ferreira}},
  \bibinfo{author}{\bibfnamefont{G.}~\bibnamefont{Giansiracusa}},
  \bibnamefont{and} \bibinfo{author}{\bibfnamefont{U.}~\bibnamefont{Lombardo}},
  \bibinfo{journal}{Phys. Lett. B} \textbf{\bibinfo{volume}{215}},
  \bibinfo{pages}{19} (\bibinfo{year}{1988}).

\bibitem[{\citenamefont{Soma and Bozek}(2006)}]{bozek06}
\bibinfo{author}{\bibfnamefont{V.}~\bibnamefont{Soma}} \bibnamefont{and}
  \bibinfo{author}{\bibfnamefont{P.}~\bibnamefont{Bozek}},
  \bibinfo{journal}{Phys. Rev. C} \textbf{\bibinfo{volume}{74}},
  \bibinfo{pages}{045809} (\bibinfo{year}{2006}).

\bibitem[{\citenamefont{Fetter and Walecka}(2003)}]{fetter}
\bibinfo{author}{\bibfnamefont{A.~L.} \bibnamefont{Fetter}} \bibnamefont{and}
  \bibinfo{author}{\bibfnamefont{J.~D.} \bibnamefont{Walecka}},
  \emph{\bibinfo{title}{Quantum Theory of Many-Particle Systems}}
  (\bibinfo{publisher}{Dover}, \bibinfo{address}{NY}, \bibinfo{year}{2003}).

\bibitem[{\citenamefont{Moshfegh and Modarres}(2005)}]{moshfegh05}
\bibinfo{author}{\bibfnamefont{H.~R.} \bibnamefont{Moshfegh}} \bibnamefont{and}
  \bibinfo{author}{\bibfnamefont{M.}~\bibnamefont{Modarres}},
  \bibinfo{journal}{Nucl. Phys. A} \textbf{\bibinfo{volume}{749}},
  \bibinfo{pages}{130} (\bibinfo{year}{2005}).

\bibitem[{\citenamefont{Kanzawa et~al.}(2007)\citenamefont{Kanzawa, Oyamatsu,
  Sumiyoshi, and Takano}}]{kanzawa07}
\bibinfo{author}{\bibfnamefont{H.}~\bibnamefont{Kanzawa}},
  \bibinfo{author}{\bibfnamefont{K.}~\bibnamefont{Oyamatsu}},
  \bibinfo{author}{\bibfnamefont{K.}~\bibnamefont{Sumiyoshi}},
  \bibnamefont{and} \bibinfo{author}{\bibfnamefont{M.}~\bibnamefont{Takano}},
  \bibinfo{journal}{Nucl. Phys. A} \textbf{\bibinfo{volume}{791}},
  \bibinfo{pages}{232} (\bibinfo{year}{2007}).

\bibitem[{\citenamefont{Mukherjee and Pandharipande}(2007)}]{mukherjee07}
\bibinfo{author}{\bibfnamefont{A.}~\bibnamefont{Mukherjee}} \bibnamefont{and}
  \bibinfo{author}{\bibfnamefont{V.~R.} \bibnamefont{Pandharipande}},
  \bibinfo{journal}{Phys. Rev. C} \textbf{\bibinfo{volume}{75}},
  \bibinfo{pages}{035802} (\bibinfo{year}{2007}).

\bibitem[{\citenamefont{Kohn and Luttinger}(1960)}]{luttinger60a}
\bibinfo{author}{\bibfnamefont{W.}~\bibnamefont{Kohn}} \bibnamefont{and}
  \bibinfo{author}{\bibfnamefont{J.~M.} \bibnamefont{Luttinger}},
  \bibinfo{journal}{Phys. Rev.} \textbf{\bibinfo{volume}{118}},
  \bibinfo{pages}{41} (\bibinfo{year}{1960}).

\bibitem[{\citenamefont{Bloch and de~Dominicis}(1958)}]{bloch58b}
\bibinfo{author}{\bibfnamefont{C.}~\bibnamefont{Bloch}} \bibnamefont{and}
  \bibinfo{author}{\bibfnamefont{C.}~\bibnamefont{de~Dominicis}},
  \bibinfo{journal}{Nucl. Phys.} \textbf{\bibinfo{volume}{7}},
  \bibinfo{pages}{459} (\bibinfo{year}{1958}).

\bibitem[{\citenamefont{Bloch and de~Dominicis}(1959{\natexlab{a}})}]{bloch59a}
\bibinfo{author}{\bibfnamefont{C.}~\bibnamefont{Bloch}} \bibnamefont{and}
  \bibinfo{author}{\bibfnamefont{C.}~\bibnamefont{de~Dominicis}},
  \bibinfo{journal}{Nucl. Phys.} \textbf{\bibinfo{volume}{10}},
  \bibinfo{pages}{181} (\bibinfo{year}{1959}{\natexlab{a}}).

\bibitem[{\citenamefont{Bloch and de~Dominicis}(1959{\natexlab{b}})}]{bloch59b}
\bibinfo{author}{\bibfnamefont{C.}~\bibnamefont{Bloch}} \bibnamefont{and}
  \bibinfo{author}{\bibfnamefont{C.}~\bibnamefont{de~Dominicis}},
  \bibinfo{journal}{Nucl. Phys.} \textbf{\bibinfo{volume}{10}},
  \bibinfo{pages}{509} (\bibinfo{year}{1959}{\natexlab{b}}).

\bibitem[{\citenamefont{terHaar and Malfliet}(1986)}]{terhaar86}
\bibinfo{author}{\bibfnamefont{B.}~\bibnamefont{terHaar}} \bibnamefont{and}
  \bibinfo{author}{\bibfnamefont{R.}~\bibnamefont{Malfliet}},
  \bibinfo{journal}{Phys. Rev. Lett.} \textbf{\bibinfo{volume}{56}},
  \bibinfo{pages}{1237} (\bibinfo{year}{1986}).

\bibitem[{\citenamefont{Huber et~al.}(1998)\citenamefont{Huber, Weber, and
  Weigel}}]{huber98}
\bibinfo{author}{\bibfnamefont{H.}~\bibnamefont{Huber}},
  \bibinfo{author}{\bibfnamefont{F.}~\bibnamefont{Weber}}, \bibnamefont{and}
  \bibinfo{author}{\bibfnamefont{M.~K.} \bibnamefont{Weigel}},
  \bibinfo{journal}{Phys. Rev. C} \textbf{\bibinfo{volume}{57}},
  \bibinfo{pages}{3484} (\bibinfo{year}{1998}).

\bibitem[{\citenamefont{Kadanoff and Baym}(1962)}]{kadanoff62}
\bibinfo{author}{\bibfnamefont{L.~P.} \bibnamefont{Kadanoff}} \bibnamefont{and}
  \bibinfo{author}{\bibfnamefont{G.}~\bibnamefont{Baym}},
  \emph{\bibinfo{title}{Quantum Statistical Mechanics}}
  (\bibinfo{publisher}{Benjamin}, \bibinfo{address}{N.Y.},
  \bibinfo{year}{1962}).

\bibitem[{\citenamefont{Mattuck}(1992)}]{mattuck}
\bibinfo{author}{\bibfnamefont{R.~D.} \bibnamefont{Mattuck}},
  \emph{\bibinfo{title}{A Guide to {F}eynman Diagrams in the Many-Body
  Problem}} (\bibinfo{publisher}{Dover}, \bibinfo{address}{NY},
  \bibinfo{year}{1992}).

\bibitem[{\citenamefont{Dickhoff and Barbieri}(2004)}]{dickhoff04}
\bibinfo{author}{\bibfnamefont{W.~H.} \bibnamefont{Dickhoff}} \bibnamefont{and}
  \bibinfo{author}{\bibfnamefont{C.}~\bibnamefont{Barbieri}},
  \bibinfo{journal}{Prog. Part. Nucl. Phys.} \textbf{\bibinfo{volume}{52}},
  \bibinfo{pages}{377} (\bibinfo{year}{2004}).

\bibitem[{\citenamefont{Schnell et~al.}(1996)\citenamefont{Schnell, Alm, and
  R{\"o}pke}}]{alm96b}
\bibinfo{author}{\bibfnamefont{A.}~\bibnamefont{Schnell}},
  \bibinfo{author}{\bibfnamefont{T.}~\bibnamefont{Alm}}, \bibnamefont{and}
  \bibinfo{author}{\bibfnamefont{G.}~\bibnamefont{R{\"o}pke}},
  \bibinfo{journal}{Phy. Lett. B} \textbf{\bibinfo{volume}{387}},
  \bibinfo{pages}{443} (\bibinfo{year}{1996}).

\bibitem[{\citenamefont{Frick and M{\"u}ther}(2003)}]{frick03}
\bibinfo{author}{\bibfnamefont{T.}~\bibnamefont{Frick}} \bibnamefont{and}
  \bibinfo{author}{\bibfnamefont{H.}~\bibnamefont{M{\"u}ther}},
  \bibinfo{journal}{Phys. Rev. C} \textbf{\bibinfo{volume}{68}},
  \bibinfo{pages}{034310} (\bibinfo{year}{2003}).

\bibitem[{\citenamefont{Frick et~al.}(2005)\citenamefont{Frick, M{\"u}ther,
  Rios, Polls, and Ramos}}]{frick05}
\bibinfo{author}{\bibfnamefont{T.}~\bibnamefont{Frick}},
  \bibinfo{author}{\bibfnamefont{H.}~\bibnamefont{M{\"u}ther}},
  \bibinfo{author}{\bibfnamefont{A.}~\bibnamefont{Rios}},
  \bibinfo{author}{\bibfnamefont{A.}~\bibnamefont{Polls}}, \bibnamefont{and}
  \bibinfo{author}{\bibfnamefont{A.}~\bibnamefont{Ramos}},
  \bibinfo{journal}{Phys. Rev. C} \textbf{\bibinfo{volume}{71}},
  \bibinfo{pages}{014313} (\bibinfo{year}{2005}).

\bibitem[{\citenamefont{Bozek and Czerski}(2002)}]{bozek1}
\bibinfo{author}{\bibfnamefont{P.}~\bibnamefont{Bozek}} \bibnamefont{and}
  \bibinfo{author}{\bibfnamefont{P.}~\bibnamefont{Czerski}},
  \bibinfo{journal}{Phys. Rev. C} \textbf{\bibinfo{volume}{66}},
  \bibinfo{pages}{027301} (\bibinfo{year}{2002}).

\bibitem[{\citenamefont{Vonderfecht et~al.}(1993)\citenamefont{Vonderfecht,
  Dickhoff, Polls, and Ramos}}]{vonder93}
\bibinfo{author}{\bibfnamefont{B.}~\bibnamefont{Vonderfecht}},
  \bibinfo{author}{\bibfnamefont{W.}~\bibnamefont{Dickhoff}},
  \bibinfo{author}{\bibfnamefont{A.}~\bibnamefont{Polls}}, \bibnamefont{and}
  \bibinfo{author}{\bibfnamefont{A.}~\bibnamefont{Ramos}},
  \bibinfo{journal}{Nucl. Phys. A} \textbf{\bibinfo{volume}{555}},
  \bibinfo{pages}{1} (\bibinfo{year}{1993}).

\bibitem[{\citenamefont{Bozek}(1999)}]{bozek99}
\bibinfo{author}{\bibfnamefont{P.}~\bibnamefont{Bozek}},
  \bibinfo{journal}{Nucl. Phys. A} \textbf{\bibinfo{volume}{657}},
  \bibinfo{pages}{187} (\bibinfo{year}{1999}).

\bibitem[{\citenamefont{M{\"u}ther and Dickhoff}(2005)}]{dickhoff05}
\bibinfo{author}{\bibfnamefont{H.}~\bibnamefont{M{\"u}ther}} \bibnamefont{and}
  \bibinfo{author}{\bibfnamefont{W.~H.} \bibnamefont{Dickhoff}},
  \bibinfo{journal}{Phys. Rev. C} \textbf{\bibinfo{volume}{72}},
  \bibinfo{pages}{054313} (\bibinfo{year}{2005}).

\bibitem[{\citenamefont{Rios et~al.}(2006{\natexlab{a}})\citenamefont{Rios,
  Polls, Ramos, and M{\"u}ther}}]{rios06}
\bibinfo{author}{\bibfnamefont{A.}~\bibnamefont{Rios}},
  \bibinfo{author}{\bibfnamefont{A.}~\bibnamefont{Polls}},
  \bibinfo{author}{\bibfnamefont{A.}~\bibnamefont{Ramos}}, \bibnamefont{and}
  \bibinfo{author}{\bibfnamefont{H.}~\bibnamefont{M{\"u}ther}},
  \bibinfo{journal}{Phys. Rev. C} \textbf{\bibinfo{volume}{74}},
  \bibinfo{pages}{054317} (\bibinfo{year}{2006}{\natexlab{a}}).

\bibitem[{\citenamefont{Rios}(2007)}]{riosphd}
\bibinfo{author}{\bibfnamefont{A.}~\bibnamefont{Rios}}, Ph.D. thesis,
  \bibinfo{school}{University of Barcelona} (\bibinfo{year}{2007}).

\bibitem[{\citenamefont{Frick}(2004)}]{frickphd}
\bibinfo{author}{\bibfnamefont{T.}~\bibnamefont{Frick}}, Ph.D. thesis,
  \bibinfo{school}{University of T{\"u}bingen} (\bibinfo{year}{2004}),
  \bibinfo{note}{advisor: Prof. H. M{\"u}ther}.

\bibitem[{\citenamefont{Baym}(1962)}]{baym62}
\bibinfo{author}{\bibfnamefont{G.}~\bibnamefont{Baym}}, \bibinfo{journal}{Phys.
  Rev.} \textbf{\bibinfo{volume}{127}}, \bibinfo{pages}{1391}
  (\bibinfo{year}{1962}).

\bibitem[{\citenamefont{Frick et~al.}(2004)\citenamefont{Frick, M{\"u}ther, and
  Polls}}]{frick04b}
\bibinfo{author}{\bibfnamefont{T.}~\bibnamefont{Frick}},
  \bibinfo{author}{\bibfnamefont{H.}~\bibnamefont{M{\"u}ther}},
  \bibnamefont{and} \bibinfo{author}{\bibfnamefont{A.}~\bibnamefont{Polls}},
  \bibinfo{journal}{Phys. Rev. C} \textbf{\bibinfo{volume}{69}},
  \bibinfo{pages}{054305} (\bibinfo{year}{2004}).

\bibitem[{\citenamefont{Rios et~al.}(2006{\natexlab{b}})\citenamefont{Rios,
  Polls, and M{\"u}ther}}]{rios06b}
\bibinfo{author}{\bibfnamefont{A.}~\bibnamefont{Rios}},
  \bibinfo{author}{\bibfnamefont{A.}~\bibnamefont{Polls}}, \bibnamefont{and}
  \bibinfo{author}{\bibfnamefont{H.}~\bibnamefont{M{\"u}ther}},
  \bibinfo{journal}{Phys. Rev. C} \textbf{\bibinfo{volume}{73}},
  \bibinfo{pages}{024305} (\bibinfo{year}{2006}{\natexlab{b}}).

\bibitem[{\citenamefont{Zuo et~al.}(2004)\citenamefont{Zuo, Li, Li, and
  Lu}}]{zuo03}
\bibinfo{author}{\bibfnamefont{W.}~\bibnamefont{Zuo}},
  \bibinfo{author}{\bibfnamefont{Z.~H.} \bibnamefont{Li}},
  \bibinfo{author}{\bibfnamefont{A.}~\bibnamefont{Li}}, \bibnamefont{and}
  \bibinfo{author}{\bibfnamefont{G.~C.} \bibnamefont{Lu}},
  \bibinfo{journal}{Phys. Rev. C} \textbf{\bibinfo{volume}{69}},
  \bibinfo{pages}{064001} (\bibinfo{year}{2004}).

\bibitem[{\citenamefont{Machleidt et~al.}(1996)\citenamefont{Machleidt,
  Sammarruca, and Song}}]{cdbonn}
\bibinfo{author}{\bibfnamefont{R.}~\bibnamefont{Machleidt}},
  \bibinfo{author}{\bibfnamefont{F.}~\bibnamefont{Sammarruca}},
  \bibnamefont{and} \bibinfo{author}{\bibfnamefont{Y.}~\bibnamefont{Song}},
  \bibinfo{journal}{Phys. Rev. C} \textbf{\bibinfo{volume}{53}},
  \bibinfo{pages}{R1483} (\bibinfo{year}{1996}).

\bibitem[{\citenamefont{Wiringa et~al.}(1995)\citenamefont{Wiringa, Stoks, and
  Schiavilla}}]{av18}
\bibinfo{author}{\bibfnamefont{R.~B.} \bibnamefont{Wiringa}},
  \bibinfo{author}{\bibfnamefont{V.~G.~J.} \bibnamefont{Stoks}},
  \bibnamefont{and}
  \bibinfo{author}{\bibfnamefont{R.}~\bibnamefont{Schiavilla}},
  \bibinfo{journal}{Phys. Rev. C} \textbf{\bibinfo{volume}{51}},
  \bibinfo{pages}{38} (\bibinfo{year}{1995}).

\bibitem[{\citenamefont{Pethick and Carneiro}(1973)}]{carneiro73}
\bibinfo{author}{\bibfnamefont{C.~J.} \bibnamefont{Pethick}} \bibnamefont{and}
  \bibinfo{author}{\bibfnamefont{G.~M.} \bibnamefont{Carneiro}},
  \bibinfo{journal}{Phys. Rev. A} \textbf{\bibinfo{volume}{7}},
  \bibinfo{pages}{304} (\bibinfo{year}{1973}).

\bibitem[{\citenamefont{Carneiro and Pethick}(1975)}]{carneiro75}
\bibinfo{author}{\bibfnamefont{G.~M.} \bibnamefont{Carneiro}} \bibnamefont{and}
  \bibinfo{author}{\bibfnamefont{C.~J.} \bibnamefont{Pethick}},
  \bibinfo{journal}{Phys. Rev. B} \textbf{\bibinfo{volume}{11}},
  \bibinfo{pages}{1106} (\bibinfo{year}{1975}).

\bibitem[{\citenamefont{Dewulf et~al.}(2003)\citenamefont{Dewulf, Dickhoff,
  Van~Neck, Stoddard, and Waroquier}}]{dewulf03}
\bibinfo{author}{\bibfnamefont{Y.}~\bibnamefont{Dewulf}},
  \bibinfo{author}{\bibfnamefont{W.~H.} \bibnamefont{Dickhoff}},
  \bibinfo{author}{\bibfnamefont{D.}~\bibnamefont{Van~Neck}},
  \bibinfo{author}{\bibfnamefont{E.~R.} \bibnamefont{Stoddard}},
  \bibnamefont{and}
  \bibinfo{author}{\bibfnamefont{M.}~\bibnamefont{Waroquier}},
  \bibinfo{journal}{Phys. Rev. Lett} \textbf{\bibinfo{volume}{90}},
  \bibinfo{pages}{152501} (\bibinfo{year}{2003}).

\bibitem[{\citenamefont{Bozek and Czerski}(2001)}]{bozek01}
\bibinfo{author}{\bibfnamefont{P.}~\bibnamefont{Bozek}} \bibnamefont{and}
  \bibinfo{author}{\bibfnamefont{P.}~\bibnamefont{Czerski}},
  \bibinfo{journal}{Euro. Jour. Phys. A} \textbf{\bibinfo{volume}{11}},
  \bibinfo{pages}{271} (\bibinfo{year}{2001}).

\bibitem[{\citenamefont{M{\"u}ther and Polls}(2000)}]{muther00}
\bibinfo{author}{\bibfnamefont{H.}~\bibnamefont{M{\"u}ther}} \bibnamefont{and}
  \bibinfo{author}{\bibfnamefont{A.}~\bibnamefont{Polls}},
  \bibinfo{journal}{Prog. Part. Nucl. Phys.} \textbf{\bibinfo{volume}{45}},
  \bibinfo{pages}{243} (\bibinfo{year}{2000}).

\bibitem[{\citenamefont{Das et~al.}(1992)\citenamefont{Das, Tripathi, and
  Sahu}}]{das92}
\bibinfo{author}{\bibfnamefont{C.}~\bibnamefont{Das}},
  \bibinfo{author}{\bibfnamefont{R.~K.} \bibnamefont{Tripathi}},
  \bibnamefont{and} \bibinfo{author}{\bibfnamefont{R.}~\bibnamefont{Sahu}},
  \bibinfo{journal}{Phys. Rev. C} \textbf{\bibinfo{volume}{45}},
  \bibinfo{pages}{2217} (\bibinfo{year}{1992}).

\bibitem[{\citenamefont{Suraud}(1998)}]{suraud}
\bibinfo{author}{\bibfnamefont{E.}~\bibnamefont{Suraud}},
  \emph{\bibinfo{title}{La Mati{\`e}re Nucl{\'e}aire: des {\'e}toiles aux
  noyaux}} (\bibinfo{publisher}{Hermann}, \bibinfo{address}{Paris},
  \bibinfo{year}{1998}).

\bibitem[{\citenamefont{Frick et~al.}(2002)\citenamefont{Frick, Gad,
  M{\"u}ther, and Czerski}}]{frick02}
\bibinfo{author}{\bibfnamefont{T.}~\bibnamefont{Frick}},
  \bibinfo{author}{\bibfnamefont{K.}~\bibnamefont{Gad}},
  \bibinfo{author}{\bibfnamefont{H.}~\bibnamefont{M{\"u}ther}},
  \bibnamefont{and} \bibinfo{author}{\bibfnamefont{P.}~\bibnamefont{Czerski}},
  \bibinfo{journal}{Phys. Rev. C} \textbf{\bibinfo{volume}{65}},
  \bibinfo{pages}{034321} (\bibinfo{year}{2002}).

\end{thebibliography}

\newpage


\begin{table}[thb]
  \begin{tabular}{ccccccc}
    \hline \hline
	Potential & Approach & $T_f$ (MeV) & $T_c$ (MeV) & $\rho_c$ (fm$^{-3}$) & $p_c$ (MeV fm$^{-3}$) & $\frac{p_c}{T_c \rho_c}$\\
    \hline
    Argonne V18 & SCGF & 9.5  & 11.6 & 0.05 & 0.08 & 0.14 \\
    \phantom    & BHF  & 13.1 & 18.1 & 0.08 & 0.40 & 0.28 \\
    CDBONN      & SCGF & 14.4 & 18.5 & 0.11 & 0.40 & 0.20 \\
	\phantom    & BHF  & 17.2 & 23.3 & 0.11 & 0.73 & 0.28 \\
    \hline \hline
  \end{tabular}

\caption{Critical properties for different approximations and NN interactions.}
\label{tab:critical}
\end{table}

\end{document}